\documentclass[aps,twocolumn,secnumarabic,balancelastpage,amsmath,amssymb,nofootinbib,floatfix,superscriptaddress]{revtex4-1}

\usepackage{amsmath}    
\usepackage{graphicx}   
\usepackage{verbatim}   
\usepackage{color}      
\usepackage{subfigure}  
\usepackage{hyperref}   
\usepackage{physics}    
\usepackage{tikz}
\usetikzlibrary{quantikz}
\usepackage[inline,shortlabels]{enumitem} 
\usepackage{hyperref} 
\hypersetup{colorlinks=true,linkcolor=blue,filecolor=magenta,urlcolor=cyan}
\usepackage{tcolorbox}

\begin{document}

\title{Quantum Blackjack\\ or\\ Can MIT Bring Down the House Again?}

\author{Joseph~X.~Lin}
\author{Joseph~A.~Formaggio}
\author{Aram~W.~Harrow}
\affiliation{Laboratory for Nuclear Science and Physics Department, Massachusetts Institute of Technology, 77 Massachusetts Ave, Cambridge, MA 02139, USA}
\author{Anand~V.~Natarajan}
\affiliation{Laboratory for Nuclear Science and Physics Department, Massachusetts Institute of Technology, 77 Massachusetts Ave, Cambridge, MA 02139, USA}
\affiliation{Institute for Quantum Information and Matter, California Institute of Technology, 1200 E. California Blvd, Pasadena, CA 91125, USA}

\date{\today}

\begin{abstract}
We examine the advantages that quantum strategies afford in communication-limited games. Inspired by the card game blackjack, we focus on cooperative, two-party sequential games in which a single classical bit of communication is allowed from the player who moves first to the player who moves second. Within this setting, we explore the usage of quantum entanglement between the players and find analytic and numerical conditions for quantum advantage over classical strategies. Using these conditions, we study a family of blackjack-type games with varying numbers of card types, and find a range of parameters where quantum advantage is achieved. Furthermore, we give an explicit quantum circuit for the strategy achieving quantum advantage.
\end{abstract}

\maketitle

\section{Introduction}

In quantum information, two-player games have provided useful perspectives on the unique power of quantum entanglement as a resource. For instance, the Clauser-Horne-Shimony-Holt (CHSH) game is an example of an operational task where quantum entanglement yields an advantage of all possible classical strategies, and analysis of CHSH---as well as more general non-local games--- has not only provided insight into foundational concepts such as Bell's inequality~\cite{cleve2004consequences}, but has also led to protocols for important tasks such as verifiable randomness generation~\cite{Col06}, key distribution~\cite{E91}, or delegated computation~\cite{RUV13}. A particularly interesting question is to understand the power of entanglement (which by itself does not permit any form of signalling) together with limited amounts of classical or quantum communication, and games play an important role here as well. In this context, a natural type of game to study is real-world communication-based strategic games. For example, bidding in the game bridge has recently been analyzed via $2\rightarrow 1$ quantum random access codes~\cite{muhammad2014quantum}. 

Inspired by the potential insight offered in studying limited communication protocols---and also drawing inspiration with our home institution's fascination with casino games~\cite{bib:blackjack}---we explore the question of whether quantum entanglement can offer a strategic advantage to win at blackjack.  In this paper, we describe how quantum entanglement can be used in blackjack, and how quantum advantages may arise. Our treatment of quantum strategies in blackjack is a special case of a communication setting that is somewhere between non-local games with no classical communication and communication complexity problems where an asymptotically growing amount of communication is used.  This area has not been heavily studied and we believe that is a promising area for finding future uses of entanglement.  Our motivation for focusing on the special case of blackjack also shows the  sort of concrete details that need to be explored in order to find settings in which Bell violations and non-locality can be used.  

The main contributions of this paper are a formalism for quantum strategies in communication-limited sequential games, a description of how optimal hyperbit strategies can be computed and realized experimentally, a concrete analysis of advantages (or lack thereof) in a small toy example, and calculations and results specific to blackjack.  

The paper is arranged as follows.  In Section~\ref{sec:blackjack-rules}, we describe the class of games we consider, as well as specify the particular assumptions and variations of blackjack used in this paper. In Section~\ref{sec:strategies}, we derive and analyze the optimal strategies for the types of games  considered in this paper. In particular, we categorize the structure and properties of the optimal strategies for three restrictions on communication: no restriction; single-bit classical communication only; and hyperbit communication. In Section~\ref{sec:hyperbit-in-practice}, we present how the optimal hyperbit strategies can actually be computed, while in Section~\ref{sec:hyperbit-proof} we give an outline for how those hyperbit strategies can be applied experimentally. In Section~\ref{sec:blackjack-games} we apply the algorithms and strategies developed herein to determine potential quantum advantage for limited communication games. We look first at low-dimensional games, which can be treated in an exact, analytic manner and then generalize these principles to larger games.  Finally, in Section~\ref{sec:backjack-results}, we describe our search for quantum advantage in concrete games of blackjack. 

\section{Blackjack and Limited Communication Games}\label{sec:blackjack-rules}
In our analysis, we consider a modified ruleset to blackjack. The purpose is to make computation and analysis simpler (and, in some cases, more feasible). Despite the apparently extreme nature of some modifications, there are specific and realistic situations that justify them.  For a more detailed description of blackjack rules, see Appendix~\ref{app:blackjack-rules}.

We consider only two players, Alice and Bob, who play cooperatively against the dealer, seeking to maximize their combined expected payout. There is no minimum or maximum bet requirement. This allows us to consider strategies in which Alice's bet is essentially zero. We do so to enforce Alice's result to be inconsequential, and therefore her action can be used simply to inform Bob's strategy. In the case in which the maximum bet far exceeds the minimum bet, this assumption is justified.

The cards are dealt as follows. The two players and the dealer each begin with a single face-up card. Each player---but not the dealer---is dealt a face-down card from the shoe. The face-up cards are public information, known to all parties, while the face-down cards are private to each player. Furthermore, the contents of the shoe prior to the deal are also public, although the order is not known. With knowledge of the public information, along with her face-down card, Alice will either hit or stand. If she hits, no matter what happens, she will stand immediately after and end her turn, since her strategy is simply to send a single bit of communication to Bob. Since her bet is taken to be inconsequential, the result of her turn is as well.

Bob uses the information available to him (including the communication from Alice) to decide whether to hit or stand. After this move, we assume the shoe is reshuffled, to a shoe containing infinitely many standard, 52-card decks. This assumption is made to simplify the analysis of the rest of the round. Bob's strategy, after his first action, can be directly calculated independently of Alice's action or private information. Such a scenario arises when the shoe becomes depleted and must be replaced. If the shoe is taken to contain many standard, 52-card decks, the rest of the round can be approximated as an infinitely-many deck shoe.

Finally, as mentioned above, Bob plays out his turn via an infinite deck strategy. The dealer then plays according to a standard strategy of hitting on a soft 17.

The payouts are then considered as follows. If Bob wins, his bet is returned to him and he wins an additional amount equal to his bet. If he loses, his bet is lost. If he ties, the bet is returned to him. 

Note finally that for all parties involved, the only allowed actions are to hit or to stand. More advanced tactics, like doubling down, splitting, or surrendering, are not considered in this exercise.

In the case that quantum entanglement is allowed, Alice and Bob initially share an arbitrary entangled state. In addition to Alice's single bit of communication, she also decides on a measurement on her half of the shared state. According to all this information, then, Bob measures his half of the state prior to the first action of his turn and uses the measurement result to decide on his action.

\subsection{Problem Statement}\label{sec:prob-statement}
Now that a particular ruleset has been described, we more generally formalize the problem. Although the discussion will be framed around our game of blackjack, more general terms will be used here and throughout the paper to refer to important concepts. In particular, the form used mirrors how non-local games are typically described.

In this paper, we consider games and strategies with the following properties. Two players, Alice and Bob, receive private information $s$ and $t$, respectively. In blackjack, this corresponds to each player's initial \textit{faced-down} card. Sequentially, each player provides a single bit response: first, Alice announces her answer $a$, followed by Bob responding his answer $b$. This corresponds to the action of \textit{hit} or \textit{stand} in blackjack. The sequential and public nature of the responses means that Bob can base his action not only on his information $t$, but also on Alice's action $a$. The goal of this game is to maximize the expected payoff function, which is constrained to only depend on $s$, $t$, and $b$. Note that we have intentionally excluded $a$, meaning Bob's action alone determines the pair's payoff. 

This formulation requires some restrictions to be placed on our version of blackjack. First, the pair's payoff cannot depend on $a$. This means that only Bob's action matters. An equivalent condition is if Bob's bet is chosen to be significantly larger than that of Alice; in the extreme case, we can make Bob's bet $1$, and Alice's $0$. Since the pair are cooperating, it is their combined payoff that matters. Second, note that their actions in any round of play is characterized only by a single bit, $a$ and $b$. For both, this means that they must play a predetermined strategy after their first hit/stand choice. In Alice's case, we simply assume she hits at most once. In this way, she only communicates one bit to Bob, and does not otherwise affect the game. In Bob's case, if he chooses $b$ to be hit, then he subsequently plays according to a predetermined strategy (much like the dealer does), without regard to Alice's sent bit $a$.

The expected payoff function then depends on the contents remaining in the shoe; this depends on $s$ and $t$, as well as the publicly available, face-up cards of all three participants. 

Finally, the task is to construct the pair's optimal strategy. Since the pair's payoff depends on Bob's action, Alice's role is to convey to Bob as much of the nature of her private information $s$ as she can. By example, this means Alice tries to communicate information about her face-down card through her action, and, if available, her correlated measurement. Bob must then utilize Alice's communication, along with his private information $t$, to optimally act.

An upper bound on the possible optimal expected payoff can be calculated assuming communication is unrestrained. Alice simply tells Bob $s$ and he has perfect information about the shoe's remaining contents. This is useful for establishing an absolute upper bound on the possible payoffs.

In blackjack and other games, however, it is sometimes against the rules to openly communicate. Thus, rather than that simple, unrestrained case, we will suppose the rules of the game restrict communication between the players. In particular, we will disallow any additional classical communication between Alice and Bob beyond those necessary for play. However, as mentioned, Alice can implicitly communicate a single bit to Bob via her public action $a$. Since the pair's payoff does not depend directly on $a$, Alice optimally uses her action to help inform Bob of her private information.

We consider one final class of strategies, which adds the use of shared quantum entanglement. In particular, we allow Alice and Bob to prepare an arbitrary quantum state before the round. Each player then receives a portion of that state, and adheres to the following protocol. Alice makes a two-outcome measurement of her state, depending on her private information. She uses that outcome as her response $a$, which she can convey to Bob. Bob then performs a two-outcome measurement of his state. As in the hyperbit model~\cite{pawlowski2012hyperbits}, which we also describe in Section~\ref{sec:strategies}, this measurement only depends on $t$, and not $a$. Finally, Bob gives his response as a function of the information he has: $t$, $a$, and his measurement.

By comparing the optimal strategies and corresponding output of these classes of strategies, we can seek if and how the addition of quantum entanglement can lead to advantages in these limited-communication games.

\begin{figure*}
    \centering
    \includegraphics[width=0.75\textwidth]{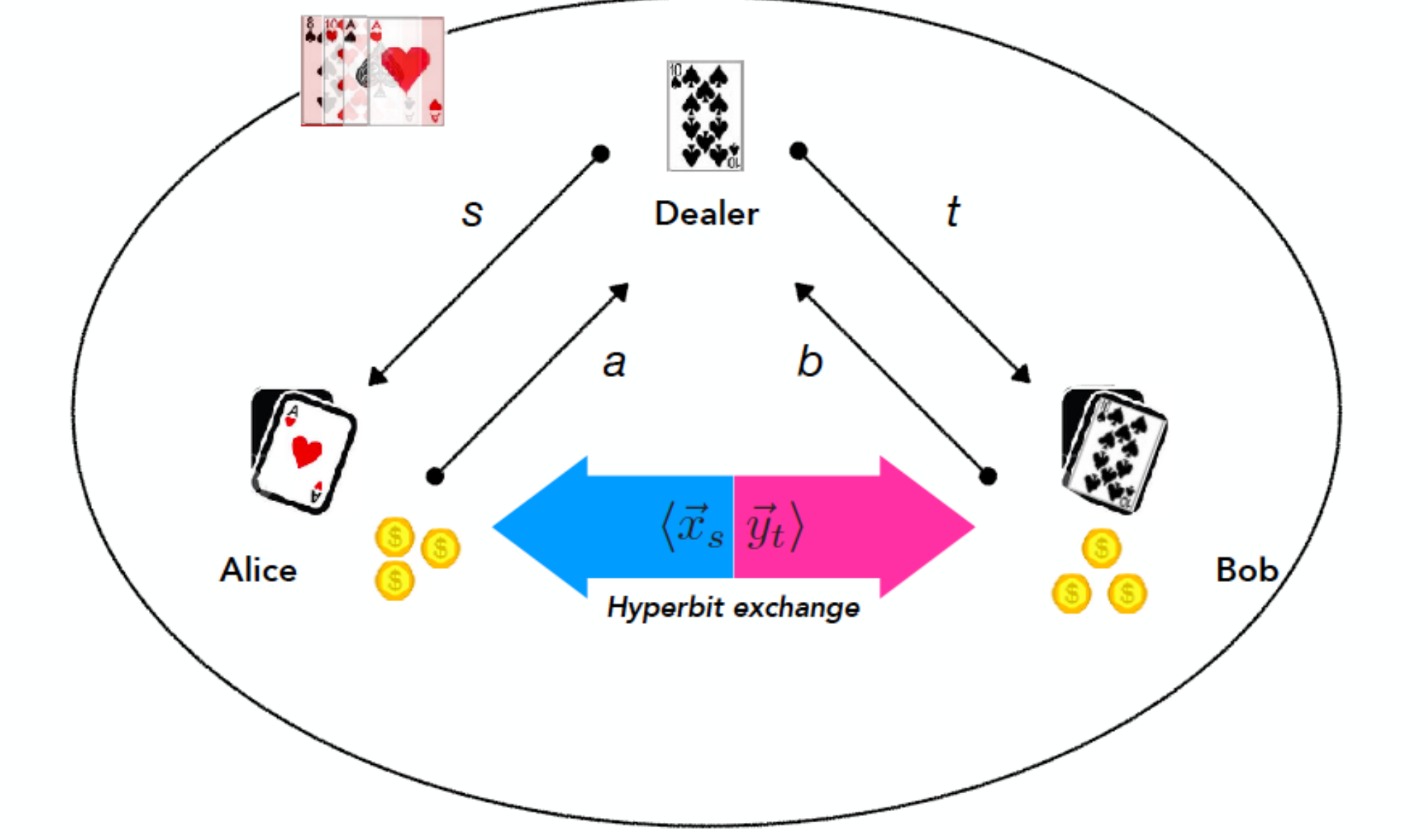}
    \caption{An illustration of the information exchange and game play considered for this paper. The dealer sends private information (face down cards) $s$ and $t$ to Alice and Bob, respectively, while Alice and Bob respond with single bit actions (hit or stand) $a$ and $b$. Bob uses public classical information (face up cards and action $a$ from Alice) to optimize his choice of action $b$. Communication can also be conveyed through measurements by Alice and Bob on shared quantum entanglement. The contents of a finite-sized shoe (4 cards in the figure) is known to the table, although the order of the cards is not.}
    \label{fig:cards_on_table}
\end{figure*}

Figure~\ref{fig:cards_on_table} provides a pictorial summary relating our notation to an example blackjack round; note again that the hyperbit notation will be presented in Section~\ref{sec:strategies}.

\subsection{Mathematical Game Model and Notation}\label{sec:game-model}
In this section, we provide a mathematical description of the types of games we previously outlined. We will chiefly describe the problem in more general terms, rather than in terms of blackjack; however, the parallels drawn in Section~\ref{sec:prob-statement} can be referred to for context and motivation.

Suppose that the pair's expected payoff is characterized by a function $V(b \lvert s, t)$. Furthermore, suppose that $s$ and $t$ are governed by the joint prior probability $\pi(s, t)$, such that they are given to Alice and Bob, respectively, with that probability. 

Next, consider the players' strategy, given a pair of private information $(s, t)$. The strategy can be characterized by the probability distribution of Bob's answer, $b$. Let that probability distribution be given by $p(b \lvert s, t)$\footnote{We have labeled Bob's two answer choices explicitly as $\pm 1$, rather than the more common choices of bits $0$ and $1$. This convention is intentionally chosen to simplify calculations, but is done so without loss of generality.}. Then, the expected payoff ${\cal P}$ is given by
\begin{equation}\label{eq:expected-payoff}
{\cal P} = \sum_{s}\sum_{t}\sum_{b \in \{\pm 1\}} \pi(s, t) V(b\lvert s, t)p(b \lvert s, t).
\end{equation}

If we let $p^{\pm}_{s, t} = p(\pm1\lvert s, t)$ and $V^{\pm}_{s, t} = V(\pm 1\lvert s, t)$, we can rearrange Equation~\ref{eq:expected-payoff} to find

\begin{align}
\label{eq:rearranged-expected-payoff}
{\cal P} = & \sum_{s, t}\pi(s, t)\Bigl( p^{+}_{s, t}V^{+}_{s, t} + (1 - p^{+}_{s, t}) V^{-}_{s, t} \Bigr) \nonumber \\ 
 = & \sum_{s, t}\pi(s, t)\biggl(\frac{(V^{+}_{s, t} - V^{-}_{s, t})(p^{+}_{s, t} - p^{-}_{s, t})}{2}  + \frac{V^{+}_{s, t} + V^{-}_{s, t}}{2} \biggr). 
\end{align}

Given this payoff function, the problem of finding the optimal solution then becomes an optimization problem over the probability distributions $p(b\lvert s, t)$. Note in particular that some terms in the right hand side of Equation~\ref{eq:rearranged-expected-payoff} do not depend on $p(b\lvert s, t)$, and are thus irrelevant to the optimization problem.

To simplify the notation, let us define two matrices, ${\bf C}$ and ${\bf S}$, such that their entries are given by
\begin{equation}
C_{st} := \pi(s, t)(V^{+}_{s, t} - V^{-}_{s, t}),
\end{equation} 
and
\begin{equation}
S_{st} := p^{+}_{s,t} - p^{-}_{s, t}.
\end{equation}

Finally, we can use Equation~\ref{eq:rearranged-expected-payoff} to write the objective function we wish to maximize as
\begin{equation}
I({\bf S}) = \sum_{s, t} C_{st}S_{st} = \langle {\bf C}, {\bf S}\rangle,
\end{equation}
where $\langle {\bf A}, {\bf B}\rangle = \text{tr}({\bf A}^T{\bf B})$ is the matrix inner product. Note that we have removed terms that do not depend on $p(b|s, t)$ and scaled the resulting terms by $2$; none of these changes affect the optimal argument resulting from the optimization problem.

The optimization problem can then be written as finding 
\begin{equation}\label{eq:optimization}
\mathbf{S}^{*} = \arg\max_{\mathbf{S}} I (\mathbf{S})
\end{equation}
which achieves the objective function value $I^{*} = I(\bf{S}^{*})$. In the rest of this text, we will refer to $I^{*}$ as the \textit{value} of the game.

Note the roles that the two matrices, ${\bf C}$ and ${\bf S}$, serve in this problem. The elements $C_{st}$ of ${\bf C}$ give an indication of how likely the private information $(s, t)$ is to be given to Alice and Bob, combined with the ``bias'' in expected payoff given to Bob answering $b = +1$ over $b = -1$. The sign of $C_{st}$ indicates how Bob would optimally answer, given perfect information, and the magnitude indicates the relative importance of correctly answering given this particular pair of private information. This matrix is a constant, determined by the properties of the game.

The elements $S_{st}$ of ${\bf S}$ give the expected value of Bob's answer, and is thus constrained by $-1\leq S_{st}\leq 1$. Additional constraints can be added, depending again on the rules of the game. For the types of games we are interested in, the constraints arise from the restrictions placed on communication between Alice and Bob. The structure of the optimal strategies, as well as the optimal payoff that these strategies can achieve, thus depend on these matrix constraints.

\section{Classical and Quantum Strategies}\label{sec:strategies}
In the unlimited communication (or perfect information) case, Alice is able to provide Bob her private information $s$ in full. It is simple enough to optimize Eq.~\ref{eq:optimization} by taking
\begin{equation}\label{eq:optimal-unlimited-S}
S^{*}_{st} = \text{sgn } C_{st}
\end{equation}
so that the maximum possible game value is given by
\begin{equation}
I^{*}_U = \sum_{s,t}\vert C_{st}\vert.
\end{equation}
While simple in nature, this result provides the upper bound to the value of games with any level of communication restriction.  Notationally, we will designate all quantities related to this class of strategies with subscript or superscript $U$.

In the case of classical communication, we are limited to a single bit, conveyed by Alice's action $a$. Most generally, for each $s$, Alice can act according to some probability distribution. Suppose Alice answers  $a = +1$ with probability $p_s$ and $a = -1$ with probability $q_s$, such that $p_s+q_s = 1$ and $0\leq p_s, q_s\leq 1$. Furthermore, suppose Bob also answers according to a probability distribution that depends on $a$ and $t$. Recall that $S_{st}$ characterizes the expected value of Bob's action $b$. Thus, it suffices to characterize Bob's individual choice by $\alpha_t$ and $\beta_t$, the expected value of $b$ given $a = +1$ and $a = -1$, respectively. In this case, $-1\leq \alpha_t, \beta_t\leq 1$. Then, the structure of ${\bf S}$ is constrained according to
\begin{equation}\label{eq:classical-S}
S_{st} = p_s\alpha_t + q_s\beta_t.
\end{equation}

From the linearity of this problem, we can derive several properties. First, suppose Bob's actions are fixed. Then,
\begin{equation}\label{eq:classical-bob-fixed}
I({\bf S}) = \sum_s\biggl( p_s\sum_t C_{st}\alpha_t + q_s\sum_t C_{st}\beta_t\biggr).
\end{equation}

For each $s$, there are then three cases to consider. If $\sum_t C_{st}\alpha_t > \sum_t C_{st}\beta_t$, then the above is optimized by taking $p_s = 1$. If, on the other hand, $\sum_t C_{st}\alpha_t < \sum_t C_{st}\beta_t$, it is optimal to take $p_s = 0$. In the case the two quantities are equal, Alice is indifferent between all strategies; in particular, however, the deterministic $p_s = 0$ and $p_s = 1$ strategies are still optimal.

If Alice's actions are fixed, then one can write the objective function as

\begin{equation}
I({\bf S}) = \sum_t\biggl(\alpha_t\sum_{s}C_{st}p_s + \beta_t\sum_s C_{st}q_s\biggr).
\end{equation}

The optimal actions are determined by fulfilling the conditions
\begin{align}
\alpha_t &= \text{sgn } \sum_s C_{st}p_s\label{eq:alpha-fixed-p},\\
\beta_t &= \text{sgn } \sum_s C_{st}q_s\label{eq:beta-fixed-p}.
\end{align}

All this is also to say that Bob will determine his action depending on $t$ and Alice's strategy. 

If we imagine the space of all possible strategies of the form in Equation~\ref{eq:classical-S}, this analysis has shown that optimal strategies for any game can be found at extremal, corner points. In particular, the constraints on all parameters are saturated, namely: $p_s = 1-q_s \in \{0, 1\}$ and $\alpha_t, \beta_t \in \{-1, 1\} $. Notationally, we will designate all quantities related to this class of strategies with subscript or superscript $C$.

We now consider the case in which Alice and Bob are still limited to a single bit of classical communication, but are allowed correlated measurements to a shared, bipartite quantum state $\rho$. In this way, Alice can convey additional information about her private question $s$ through quantum entanglement. 

There is, admittedly, a large space of possible quantum strategies to consider. To make analysis more tractable, we consider only the strategies which utilize the entanglement and communication according a hyperbit protocol~\cite{pawlowski2012hyperbits}. Finding scenarios where hyperbit strategies are advantageous over classically restricted strategies would indicate regimes in which quantum advantages exist. However, it is also important to note that the absence of hyperbit advantages does not rule out quantum advantages arising from even more general quantum strategies.

Within the hyperbit model, Alice can prepare a hyperbit according to a vector $\vec{x}_s$, which Bob can subsequently measure according to a vector $\vec{y}_t$. The expected value of Bob's hyperbit measurement is determined by the quantity $\vec{x}_s\cdot\vec{y}_t$. However, in choosing his answer $b$, Bob can choose, according to $t$, to probabilistically use the hyperbit measurement, flip the hyperbit measurement, or default to some deterministic answer. This amounts to adding an offset $\gamma_t$ to a scaled version of $\vec{x}_s\cdot\vec{y}_t$. Note the scaling need not be made explicit, as it can be absorbed into the choice of $\vec{y}_t$.

The mathematical description of this set of strategies can thus be given as 
\begin{equation}\label{eq:hyperbit-simplified-S}
S_{st} = \gamma_{t} + \vec{x}_{s}\cdot \vec{y}_{t}.
\end{equation}

To fully categorize these strategies, restrictions need to be placed on the parameters. The overall restriction that must be satisfied is that $\lvert S_{st}\rvert\leq 1$ for all $s$ and $t$, and all valid choices of parameters.

 Suppose the space of Alice's strategy is restricted such that $\lVert\vec{x}_s\rVert\leq c$ for all $s$, for some fixed value $c>0$. We then note that
 \begin{equation}
 \lvert S_{st}\rvert = \lvert  \gamma_{t} + \vec{x}_{s}\cdot \vec{y}_{t}\rvert \leq \lvert \gamma_{t}\rvert + \lvert\vec{x}_s\cdot\vec{y}_t\rvert\leq \lvert \gamma_{t}\rvert +c \lVert\vec{y}_t\rVert.
 \end{equation}
The above inequalities are tight; it is thus necessary for Bob's strategy to be restricted by
\begin{equation}
\lvert \gamma_{t}\rvert +c \lVert\vec{y}_t\rVert\leq 1.
\end{equation}

It turns out that it is most convenient to follow the convention used in~\cite{pawlowski2012hyperbits}, and choose $c = 1$; this allows for the hyperbit vectors to be directly interpretable as $\pm 1$-outcome measurement operators. 

In summary, the sole restrictions that need to be followed are that
\begin{equation}
\lVert \vec{x}_s\rVert \leq 1
\end{equation}
for all $s$, and
and
\begin{equation}
\lvert\gamma_t\rvert + \lVert\vec{y}_t\rVert\leq 1
\end{equation}
for all $t$. 

As with the classically restricted case, the problem can now be rephrased as an optimization problem. In particular, the strategy matrix given by the form in Equation~\ref{eq:hyperbit-simplified-S} allows the objective function to be written as

\begin{equation}
I({\bf S}) = \sum_{s,t}C_{st}(\gamma_{t} + \vec{x}_{s}\cdot \vec{y}_{t}). \end{equation}

As was the case with the classical communication only analysis above, we can determine that the aforementioned inequality constraints are saturated to be equalities in the optimal solution.

First, suppose Bob's actions are fixed. Regrouping the objective function, we find that
\begin{equation}
I({\bf S}) = \sum_s\biggl(\sum_t C_{st}\gamma_t + \vec{x}_s\cdot\sum_t C_{st}\vec{y}_t\biggr).
\end{equation}
It is thus optimal to choose 
\begin{equation}\label{eq:alice-strat-given-bob}
\vec{x}_s = \biggl(\sum_t C_{st}\vec{y}_t\biggr)\bigg/\biggl\Vert\sum_t C_{st}\vec{y}_t\biggr\Vert
\end{equation}
i.e. choose $\vec{x}_s$ to be as long as possible and in the direction of $\sum_t C_{st}\vec{y}_t$. The general condition is thus that $\Vert\vec{x}_s\Vert = 1$. Essentially, Alice should ``send as much information as possible,'' which is done by sending a unit vector.

Next, suppose Alice's actions are fixed. Again, regrouping the objective function gives
\begin{equation}
I({\bf S}) = \sum_t\biggl(\gamma_t\sum_s C_{st} + \vec{y}_t\cdot\sum_s C_{st}\vec{x}_s\biggr).
\end{equation}
Before applying any restrictions, we consider independently optimizing each of the two terms in the parentheses. Recall that the $\gamma$'s and $\vec{y}$'s are only mutually constrained by their magnitudes. In particular, we choose the sign of the $\gamma$'s and the direction of the $\vec{y}$'s independently. For the left term, note that we take 
\begin{equation}\label{eq:gamma-sign}
\text{sgn }\gamma_t = \text{sgn }\Bigl(\sum_s C_{st}\Bigr)
\end{equation} 
while, for the right term, we take $\vec{y}_t$ to be in the direction of 
\begin{equation}
\sum_s C_{st}\vec{x}_s.
\end{equation}
In this way, the objective function becomes
\begin{equation}
\sum_t\biggl(\vert\gamma_t\vert\biggl\vert\sum_s C_{st}\biggr\vert + \Vert\vec{y}_t\Vert\biggl\Vert\sum_s C_{st}\vec{x}_s\biggr\Vert\biggr)
\end{equation}

There are then two cases to consider. If $\biggl\vert\sum_s C_{st}\biggr\vert  > \biggl\Vert\sum_s C_{st}\vec{x}_s\biggr\Vert$, then Bob should choose $\vert\gamma_t\vert = 1$ (so that $\Vert\vec{y}_t\Vert = 0$); otherwise, Bob should choose $\Vert\vec{y}_t\Vert = 1$ (so that $\vert\gamma_t\vert = 0$). This is all in very much the same vein as the analysis of Equation~\ref{eq:classical-bob-fixed}. 

Thus, taking note of all these properties, we can conclude that for each $(s, t)$, we have either
\begin{equation}\label{eq:default-option}
S^{*}_{st} = \gamma_t
\end{equation}
or
\begin{equation}\label{eq:hyperbit-option}
S^{*}_{st} = \vec{x}_s\cdot \vec{y}_t.
\end{equation}
This result can be interpreted as follows. For particular values of $t$, Bob can choose to ignore all communication and default to a deterministic option; this corresponds to Equation~\ref{eq:default-option}. In the other case, given in Equation~\ref{eq:hyperbit-option}, both Alice and Bob will act according to a hyperbit model. As given by results from Tsirelson~\cite{tsirel1987quantum}, there exist a bipartite state $\rho$ and measurement operators $\hat{A}_s$ and $\hat{B}_t$ for Alice and Bob, respectively, such that
\begin{equation}\label{eq:tsirelson-condition}
\vec{x}_s\cdot\vec{y}_t = \text{Tr}(\hat{\bf A}_s\otimes\hat{\bf B}_t\rho) = \langle {\bf AB}\rangle.
\end{equation}
If Alice sends Bob her measurement result ${\bf A}$, Bob can take his measurement result ${\bf B}$ and respond with his answer as $b = {\bf AB}$. This, by construction, has expected value given by Equation~\ref{eq:tsirelson-condition}.

To calculate the exact optimal strategy matrix $S^{*}$, numerical strategies must be employed. The discussion in this section greatly simplifies the problem and limits the cases that need to be considered, making the optimization feasible. Notationally, we will designate all quantities related to this class of strategies with subscript or superscript $H$.

\section{Computing Optimal Hyperbit Strategies}\label{sec:hyperbit-in-practice}
To compute hyperbit strategies, it is most straightforward to enumerate certain discrete choices, and take the optimal result from considering the subcases. First, we fix Bob's default moves. Since $\gamma_t$ is optimally either $0$ or $\text{sgn }\Bigl(\sum_s C_{st}\Bigr)$ (from Equation~\ref{eq:gamma-sign}), we can explicitly enumerate the finitely many vectors $\vec{\gamma}$ of default strategies to consider. Next, for every one of these $\vec{\gamma}$ vectors, we can perform the following procedure.

Numerically, for a fixed set of game parameters ${\bf C}$, and default Bob strategies $\vec{\gamma}$, we attempt to compute the optimal strategy matrix ${\bf S}$. Recall then that whenever $\gamma_t = 0$, then optimally $\lVert \vec{y}_t\rVert = 1$, and whenever $\lvert \gamma_t\rvert = 1$, then $\lVert \vec{y}_t\rVert = 0$ is required. The objective function to maximize becomes
\begin{equation}\label{eq:sdp-obj}
I({\bf S}) = \sum_{\substack{t \\ \gamma_t \neq 0}}\biggl\vert\sum_s C_{st}\biggr\vert + \sum_{s}\sum_{\substack{t \\ \gamma_t = 0}}C_{st}\vec{x}_s\cdot\vec{y}_t.
\end{equation}

Only the second set of double-summations needs to be maximized. This can be done via a semi-definite program as follows. Let ${\bf C}^{\prime}$ be some columns of ${\bf C}$, such that column $t$ of ${\bf C}$ is included if and only if $\gamma_t = 0$. Let ${\bf X}$ contain as columns all the vectors $\vec{x}_s$, and let ${\bf Y}$ contain as columns all the \textit{nonzero} vectors $\vec{y}_t$. Let ${\bf Z} = \begin{pmatrix} {\bf X} & {\bf Y}\end{pmatrix}$ then be the concatenation of ${\bf X}$ and ${\bf Y}$. Now, consider the matrix given by ${\bf G}= {\bf Z}^T{\bf Z}$. As a block schematic, it can be written as
\begin{equation}\label{eq:block}
{\bf G}= \left(
\begin{array}{c|c}
{\bf X}^T {\bf X} & {\bf X}^T {\bf Y} \\
\hline
{\bf Y}^T {\bf X} & {\bf Y}^T {\bf Y}
\end{array}
\right) = 
\left(
\begin{array}{c|c}
\vec{x}_i\cdot \vec{x}_j & \vec{x}_i\cdot \vec{y}_j \\
\hline
\vec{y}_i\cdot \vec{x}_j & \vec{y}_i\cdot \vec{y}_j
\end{array}
\right)
\end{equation}
\noindent where the dot products indicate the elements at row $i$ and column $j$ within the block they reside. Note that, by definition, this is a Gramian matrix and thus is positive semi-definite. Therefore, our optimization problem can be written as a semi-definite program, and the matrix ${\bf G}$ can be solved for numerically. The primal problem can be written as
\begin{equation}\label{eq:sdp}
\begin{aligned}
& \underset{\bf G}{\text{maximize}}
& & \langle {\bf D}, {\bf G}\rangle \\
& \text{subject to}
& & \text{diag}({\bf G}) = e, \\
&&& {\bf G}\succeq 0
\end{aligned}
\end{equation}
where
\begin{equation}
{\bf D} = \frac{1}{2}\begin{pmatrix}
{\bf 0} & {\bf C}^{\prime}\\
{\bf C}^{\prime T} & {\bf 0}
\end{pmatrix}
\end{equation}
and $e = \begin{pmatrix} 1 & 1 & \cdots & 1\end{pmatrix}^T$.
Efficient numerical methods are known that can solve this problem, including PICOS~\cite{picos} (our SDP solver of choice) and CVX~\cite{grant2008cvx}. 

With this known, the calculated values in ${\bf G}$ can be substituted into Equation~\ref{eq:sdp-obj}. By solving the corresponding SDP for every possible vector of values $\vec{\gamma}$, the optimal strategy that maximizes Equation~\ref{eq:sdp-obj} can be established.

Once the optimal ${\bf G}$ is found, the optimal hyperbit vectors $\vec{x}_s$ and $\vec{y}_t$ can be determined. First, suppose that the $\vec{x}_s$ vectors are indexed by $s\in\{1, 2, \ldots, m\}$ and the nonzero $\vec{y}_t$ vectors are indexed by $t\in\{1, 2, \ldots, n\}$. ${\bf G}$ will then have dimensions $(m + n)\times (m + n)$. Next, note that ${\bf G}$ can be factorized via the Cholesky decomposition to give
\begin{equation}
{\bf G}= {\bf U}^T{\bf U}
\end{equation}
for some upper triangular matrix ${\bf U}$. While we could naively take the first $m$ columns of ${\bf U}$ to be the $\vec{x}$'s and the rest to be $\vec{y}$'s, there is a strategy that can be used to reduce the dimensionality of these $(m+n)$-component vectors. This is due to two facts. First is that the lower $n$ components in each of the first $m$ columns of ${\bf U}$ is $0$. Second is the fact that, from our discussion in Section~\ref{sec:strategies}, for $\gamma_t = 0$, a fixed choice for the $\vec{x}_s$ vectors uniquely determines the optimal set of $\vec{y}_t$ vectors (and vice versa).

Suppose $m \leq n$. Recall from Section~\ref{sec:strategies} that if the $\vec{x}_s$ vectors are optimally fixed and $\gamma_t = 0$ for all $t$, then we can choose
\begin{equation}\label{eq:y-from-x}
\vec{y}_t  = \biggl(\sum_s C_{st}\vec{x}_s\biggr)\bigg/\biggl\Vert\sum_s C_{st}\vec{x}_s\biggr\Vert.
\end{equation}
Rather than Cholesky factorizing the entire Gramian matrix ${\bf G}$, then, it suffices to Cholesky factorize only the ${\bf X}^T{\bf X}$ block of ${\bf G}$ (recall the block form in Equation~\ref{eq:block}). The columns of the resulting triangular matrix, then, are only $m$-component vectors, which can be taken directly to be $\vec{x}_s$. Equation~\ref{eq:y-from-x} can be used to calculate $\vec{y}_t$. 

If instead we had $n < m$, a similar procedure can be performed. However, first the $\vec{y}_t$ vectors are determined by Cholesky factorizing the ${\bf Y}^T{\bf Y}$ block, to get $n$-component vectors. Then, $\vec{x}_s$ can be determined using Equation~\ref{eq:alice-strat-given-bob}. 

This procedure thus reduces the dimension of the vectors from $m + n$ to 
\begin{equation}
d \equiv \text{min}(m, n).
\end{equation}
As will be seen, this allows for the size of the shared state to be reduced as well.

It should be pointed out that in arbitrary hyperbit strategies, which may be non-optimal, the dimensionality of the vectors can only generally reduced to $\text{min}(m,n)+2$~\cite{tsirel1987quantum}. Our analysis of the optimal hyperbit strategies, however, allows us to use the fact that the $\vec{y}_t$ vectors can be written as a sum of the $\vec{x}_s$ vectors, and vice versa, to further reduce the dimensionality.

\section{Implementing the Hyperbits Algorithm on Quantum Hardware}\label{sec:hyperbit-proof}
Here, we consider how the results of the aforementioned theoretical and computational analyses can be used in an experimental setting. In particular, we consider how to derive the exact quantum operations (gates) and measurements consistent with the theoretical hyperbit strategies. We achieve this goal in several steps, with each step transitioning to a lower level of abstraction from the previous.

\subsection{Deriving State and Measurements from Hyperbit Vectors}\label{sec:hyperbit-to-measurement}
After determining the optimal hyperbit vectors (Section~\ref{sec:hyperbit-in-practice}) $\vec{x}_s, \vec{y}_t\in\mathbb{R}^{d}$, we must specify the actual state and unitary measurements that are made. Following the Tsirelson characterization of 2-player XOR games~\cite{vidick2014tsirelson}, let 
\begin{equation}
L \equiv \biggl\lceil \frac{d}{2}\biggr\rceil
\end{equation}
where $L$ represents the number of qubits each player receives; the dimension of the shared state is therefore $2^{2L}$. Then, let $\rho = \ket{\Psi}\bra{\Psi}$, for the maximally entangled state
\begin{equation}
\ket{\Psi} = \frac{1}{2^{L/2}}\sum_{i = 0}^{2^L - 1}\ket{i}\ket{i}.
\end{equation}
Note that each state $\ket{i}$ is a $L$-qubit register.

Next, for $1\leq i \leq L$, we define the operators
\begin{align}
T_{2i-1} &= X_i\prod_{j = 1}^{i - 1} Z_j\\
T_{2i} &= Y_i\prod_{j = 1}^{i - 1} Z_j
\end{align}
such that $X_k$, $Y_k$, and $Z_k$ indicate the action of the $X$, $Y$, and $Z$ Pauli matrices, respectively, on qubit $k$ of a register, and identity matrices on all other qubits. These operators are chosen to fulfill the anti-commutation relation
\begin{equation}
\{T_i, T_j\} = 2\delta_{ij}\mathbb{I}.
\end{equation}  
This property allows us to define our measurement operators, in terms of these $T$ operators, as
\begin{align}
\hat{\bf A}_s &= \sum_{i = 1}^{d}(\vec{x}_s)_i T_i \label{eq:alice-measurement}\\
\hat{\bf B}_t &= \sum_{i = 1}^{d}(\vec{y}_t)_i T_i^{\text{T}} \label{eq:bob-measurement}
\end{align}
such that
\begin{equation}
\bra{\Psi}\hat{\bf A}_s\otimes\hat{\bf B}_t\ket{\Psi} = \vec{x}_s\cdot\vec{y}_t.
\end{equation}
Additionally
\begin{equation}
\hat{\bf A}_s^2 = \|\vec x_s\|^2 \mathbb{I} \quad\text{and}\quad
  \hat{\bf B}_t^2 = \|\vec y_t\|^2 \mathbb{I}.
\label{eq:AB2}\end{equation}

Equation~\ref{eq:AB2} implies that $\hat{\bf A}_s$ and $\hat{\bf B}_t$ can be realized by $\pm 1$-valued measurements as long as $\|\vec x_s\|\leq 1$ and $\|\vec y_t\|\leq 1$; as discussed in Section~\ref{sec:strategies}, $\vec x_s$ and $\vec y_t$ are unit vectors in cases relevant to us. If Alice and Bob measure their part of the state according to $\hat{\bf A}_s$ and $\hat{\bf B}_t$, respectively, Alice sends her measurement result to Bob, and Bob acts according to the product of their two measurements, then his expected value action is given by $\vec{x}_s\cdot\vec{y}_t$, consistent with the hyperbit strategies. 
\subsection{Decomposing Measurements into Single- and Two-Qubit Gates}\label{sec:measurements-to-gates}
In the previous section, we were able to explicitly write the unitary matrices corresponding to the measurement operators. In this section, we explore one possible implementation of those measurements.

As written, the operators have dimension $2^L\times 2^L$ and are $L$-qubit gates. Arbitrary $L$ qubit operations are in general difficult to implement; however, the particular structure of the measurement operators suggests a decomposition to more feasible operations, such as single- and two-qubit gates.

First, note that the form of each operator can be written as
\begin{equation}\label{eq:measurement-string}
c_1 X_1 + c_2 Y_1 + c_3 X_2 Z_1 + c_4 Y_2 Z_1 + c_5 X_3Z_2Z_1 + c_6 Y_3Z_2Z_1 + \cdots
\end{equation}
This is immediately obvious from the form of $\hat{A}_s$ given in Equation~\ref{eq:alice-measurement}; we may simply take $c_j = (\vec{x}_s)_j$. As a special computational note, it is useful to see that
\begin{align}
T_{2i-1}^{\text{T}} &= T_{2i - 1}\\
T_{2i}^{\text{T}} &= -T_{2i}
\end{align}
since $X^{\text{T}} = X$, $Y^{\text{T}} = -Y$, and $Z^{\text{T}} = Z$. Given this fact, we see that the form of $\hat{B}_t$ given in Equation~\ref{eq:bob-measurement} can also be written in the form of Equation~\ref{eq:measurement-string} by taking $c_{2i-1} = (\vec{y}_s)_{2i-1}$ and $c_{2i} = -(\vec{y}_s)_{2i}$. 

Next, we seek to use single qubit $Z$ rotations on the state, so that the measurement operator no longer contains $Y$ Pauli's. For example, suppose a rotation of the form $e^{i\theta_1 Z_1/2}$ were applied to the state. The measurement operator then becomes
\begin{equation}
e^{i\theta_1 Z_1/2}(c_1 X_1 + c_2 Y_1 + c_2 X_2 Z_1 + c_3 Y_2 Z_1 + \cdots)e^{-i\theta_1 Z_1/2}.
\end{equation}
Aside from the first two terms of the measurement string, all other terms commute with the rotation. Thus, we can focus on the effect on the first two terms:
\begin{equation}\label{eq:x-y-rotated}
\begin{split}
e^{i\theta_1 Z_1/2}&(c_1 X_1 + c_2 Y_1 )e^{-i\theta_1 Z_1/2} \\
&= c_1(\cos\theta_1 X_1 - \sin\theta_1 Y_1) \\
&\quad+ c_2(\cos\theta_1 Y_1 + \sin\theta_1 X_1) \\
&= (c_1\cos\theta_1 + c_2\sin\theta_1)X_1 \\
&\quad+ (-c_1\sin\theta_1 + c_2\cos\theta_1)Y_1.
\end{split}
\end{equation}
To eliminate $Y_1$, we choose
\begin{equation}
\theta_1 = \arctan (c_2/c_1)
\end{equation}
so that Equation~\ref{eq:x-y-rotated} becomes
\begin{equation}
\sqrt{c_1^2 + c_2^2} X_1.
\end{equation}
Thus, if we apply the rotations $e^{i\theta_jZ_j/2}$, with $\theta_j = \arctan(c_{2j}/c_{2j-1})$, for all $1\leq j\leq L$, the resulting measurement becomes
\begin{equation}
c_1^{\prime} X_1 + c_2^{\prime} X_2Z_1 + c_3^{\prime} X_3Z_2Z_1  + \cdots + c_{N}^{\prime} X_LZ_{L-1}\cdots Z_1
\end{equation}
with $c_j^{\prime} = \sqrt{c_{2j-1}^2 + c_{2j}^2}$. 

Now, using two-qubit rotations, we seek to reduce the measurement to just an $X$ measurement on qubit $1$. To motivate this, consider the \textit{last} two terms in the sum. Suppose we applied the rotation $e^{i\phi_L A_L/2}$, where $A_L = -X_LY_{L-1}$. Because this two-qubit rotation only acts on qubits $L-1$ and $L$, only the final two terms are affected. We can note its affect by isolating the terms affecting those qubits:
\begin{equation}\label{eq:zx-x-rotated}
\begin{split}
e^{i\phi_LA_L/2}&(c^{\prime}_{L-1} X_{L-1} + c^{\prime}_L X_LZ_{L-1})e^{-i\phi_LA_L/2} \\ &= c^{\prime}_{L-1}(\cos\phi_{L} X_{L-1} - \sin\phi_{L}X_LZ_{L-1}) \\
&\quad+ c^{\prime}_{L}(\cos\phi_{L} X_LZ_{L-1} + \sin\phi_{L}X_{L-1})
\\ &=(c^{\prime}_{L-1}\cos\phi_{L} + c^{\prime}_{L}\sin\phi_{L}) X_{L-1} \\
&\quad+  (-c^{\prime}_{L-1}\sin\phi_L + c^{\prime}_{L}\cos\phi_L)X_LZ_{L-1}.
\end{split}
\end{equation}
To eliminate the $X_NZ_{N-1}$ term, we choose
\begin{equation}
\begin{split}
\phi_N &= \arctan(c^{\prime}_N/c^{\prime}_{N-1})\\
&= \arctan(\sqrt{c^2_{2N-1}+ c^2_{2N}}/\sqrt{c^2_{2N-3} + c^2_{2N-2}}).
\end{split}
\end{equation}
so that Equation~\ref{eq:zx-x-rotated} becomes
\begin{equation}
\sqrt{c^{\prime 2}_{L-1} + c^{\prime 2}_L}X_{L-1} = \sqrt{c^2_{2L-3} + c^2_{2L-2}+c^2_{2L-1}+ c^2_{2L}}X_{L-1}.
\end{equation}
Thus, if we apply two-qubit rotations of the form $e^{i\phi_jA_j/2}$, for $A_j = X_jY_{j-1}$ and $\phi_j = \arctan (\sqrt{\sum_{k = j}^L (c_{2k-1}^2 + c_{2k}^2)} / \sqrt{c_{2(j-1)-1}^2 + c_{2(j-1)}^2})$, in descending order from $j = L$ to $j = 2$, the resulting measurement becomes
\begin{equation}
\sqrt{\sum_{k = 1}^L (c_{2k-1}^2 + c_{2k}^2)} X_1
\end{equation}
which is a trivial single qubit measurement. Note that since the vectors $\vec{x}_s$ and $\vec{y}_t$ are unit vectors, the measurement is exactly just $X_1$. 

Note that the form of two-qubit rotations still appears nontrivial. To explicitly show its decomposition into simpler gates, we note that
\begin{equation}
A_j = X_j Y_{j-1} = UZ_{j}U^{\dagger}
\end{equation}
for $U = S_j \text{CNOT}_{j(j-1)}H_j$. Note here that $S_j$ is not related to the strategy matrix, but rather is the phase gate $\begin{pmatrix}
1 & 0\\ 0 & i
\end{pmatrix}$ applied to qubit $j$; $\text{CNOT}_{j(j-1)}$ is the controlled-not gate, with control qubit $j$ and target qubit $j-1$; and $H_j$ is the Hadamard gate $\dfrac{1}{\sqrt{2}}\begin{pmatrix} 1 & 1 \\ 1 & -1\end{pmatrix}$, applied to qubit $j$. 

The rotation itself can then be written as
\begin{equation}\label{eq:decomp}
e^{i\phi_jA_j/2} = Ue^{i\phi_j Z_j/2}U^{\dagger}.
\end{equation}

To summarize, the steps are:
\begin{tcolorbox}
\begin{enumerate}[label=\textit{\underline{Step \arabic*:}~}]
\item For all $1\leq j \leq L$, apply single qubit rotations $e^{i\theta_jZ_j/2}$ for $\theta_j = \arctan{(c_{2j}/c_{2j-1})}$
	\item In descending order from $j = L$ to $j=2$, apply two qubit rotations $e^{i\phi_jA_j/2}$ for $\phi_j$ equal to
	\begin{equation}
	    \phi_j \equiv \arctan (\frac{\sqrt{\sum_{k = j}^L (c_{2k-1}^2 + c_{2k}^2)} }{ \sqrt{c_{2(j-1)-1}^2 + c_{2(j-1)}^2}}).
	\end{equation} Use the decomposition given in Equation~\ref{eq:decomp} to further simplify gates.
	\item Measure the $X_{1}$ expectation value of the rotated state.
\end{enumerate}
\end{tcolorbox}

Through these steps, we have provided a method for decomposing each player's measurement into a series of elementary single qubit gates and CNOTs. It should be noted, of course, that this is not the only gate decomposition for these measurements, nor should it be considered the optimal or most efficient one. Such considerations will depend on the specifics of the device or qubits used and the gateset available. The advantage of the decomposition given here is that they are in terms of simple, single- and two-qubit gates. Furthermore, the two-qubit gates are only applied between adjacently labeled qubits, allowing for this specific gate decomposition to be applied on linear or ring architectures with nearest-neighbor connectivity. 
\section{Quantum Advantage for Limited
Communication Games}\label{sec:blackjack-games}
In this section, we consider games of varying dimensionality, i.e. the dimensions of the coefficient matrix ${\bf C}$. We will look specifically at low-dimensional games, which can be treated in an exact, analytical manner. We then generalize principles to larger games.

Let the dimensionality of ${\bf C}$ be $M\times N$. The interpretation of this is that Alice's private information, $s$, takes on one of $M$ distinct values, and Bob's private information, $t$, takes on one of $N$.

\subsection{Trivially Small or Simple Games}
When $M$ or $N$ are small, or the structure of ${\bf C}$ is especially simple, the classical strategies may be sufficient to satisfy the maximal possible payout, i.e. the payout in the unlimited information case. It is then of no use to consider quantum strategies, since they cannot possibly perform better.

We know, by virtue of the restrictions on communication, that $I^{*}_C\leq I^{*}_H\leq I^{*}_U$. Consequently, if $I^{*}_C = I^{*}_U$, then hyperbit strategies (and, indeed, any quantum strategies) can afford no advantage over classical strategies. In particular, note that the optimal classical strategy essentially amounts to Bob having two possible deterministic strategies, selected by Alice's communicated bit $a$. This corresponds to a strategy matrix ${\bf S}$ with at most $2$ distinct rows, filled with entries equal to $\pm 1$. The rows for which $p_s = 1$ are given by the $\alpha_t$ values, and the rows for which $p_s = 0$ correspond to the $\beta_t$ values.

Let $\text{sgn } {\bf C}$ be the matrix with each element replaced with its sign; we will refer
to this matrix as the \textit{sign matrix}. Recall from
Equation~\ref{eq:optimal-unlimited-S} that the best possible strategy, with no
communication restrictions, is given by the sign matrix itself. Thus, if the sign matrix has two or fewer distinct rows, the condition $I^{*}_C = I^{*}_U$ will be achieved.

In terms of the dimensions of ${\bf C}$, then, quantum advantage can only arise when $M > 2$ and $N > 1$. The former condition is because for there to be more than $2$ distinct rows, there must obviously be more than $2$ total rows; the latter is due to the fact that there can be $2^N$ possible row patterns.

\subsection{Reduction and Transformation of Games}\label{sec:red-and-trans}
The analysis of a given game can be reduced and transformed into others, reducing the space of possible games that must be analyzed. In particular, the sign matrix again affords utility in determining valid and useful reductions or transformations.

In this context, reduction refers to the reduction of dimensionality, in particular the number of relevant columns. If column $t$ of ${\bf C}$ is all the same sign, it can be altogether ignored in the analysis. This is because $\alpha_t, \beta_t$ and $\gamma_t$ can all be taken to be that sign value, independently of the other parameters. This allows both the classical and quantum strategies to optimize over that specific column in a trivial manner, and match the optimal, unrestricted communication strategy.

Thus, when considering games with particular sign matrices, only those with sign matrices containing inhomogeneous columns need to be considered.

Another strategy that can be employed are transformations of games, which is especially useful in converting games with a particular sign matrix into another. Two trivial transformations are permuting the rows and columns of ${\bf C}$; the problem can be solved for the permuted matrix, and then the optimal strategy for the original matrix can be determined by (un)permuting the solved ${\bf S}$ matrix. In terms of the problem statement, this simply refers to a relabeling of the private information.

One less trivial operation is the negation (i.e. sign flip) of an entire column of ${\bf C}$. The corresponding strategy matrix can be found by negating all of Bob's parameters (i.e. $\alpha, \beta, \gamma$ and $\vec{y}$) for that column. This corresponds to Bob remapping his response for a given choice of private information (e.g. Bob flips his hyperbit measurement outcome). 

\subsection{$3\times 2$ Dimensional Games}\label{sec:3x2-games}
The smallest possible non-trivial game one can consider is one where $(M, N) = (3,2)$. Using the transformation strategies outlined in Section~\ref{sec:red-and-trans}, it suffices to only consider games with the sign matrix
\begin{equation}\label{eq:sign-matrix}
\text{sgn } {\bf C} = \begin{pmatrix}
	1 & 1\\
	1 & -1\\
	-1 & -1
\end{pmatrix}
\end{equation}
as all others can be transformed to this one.

Before considering quantum strategies, we first characterize the optimal classical strategies. All possible classical strategies can be enumerated, and it can be confirmed that at least one of five possible strategies must be optimal. These five strategies match the optimal unrestricted communication strategy in all entries except for
\begin{enumerate*}[1)]
	\item $S_{12}$,
	\item $S_{21}$,
	\item $S_{22}$,
	\item $S_{31}$, and
	\item both $S_{11}$ and $S_{32}$.
\end{enumerate*}
These strategies then give $I_C(S)$ that is less than $I^{*}_U$ by an amount $2\delta$, for $\delta$ equal to
\begin{enumerate*}[1)]
\item $\lvert C_{12}\rvert$,
\item $\lvert C_{21}\rvert$,
\item $\lvert C_{22}\rvert$,
\item $\lvert C_{31}\rvert$, and
\item $\lvert C_{11}\rvert + \lvert C_{32}\rvert$
\end{enumerate*}
respectively. The optimal classical strategy is determined by which of these values is the \textit{smallest}; if the number is $\delta^{*}$, then the classical strategy game value is given by $I^{*}_C = I^{*}_U - 2\delta^{*}$. 

With a sense of the classical strategies, we shift our attention to hyperbit strategies. To start, we will only explore the hyperbit strategies for which $\gamma_t = 0$ for all $t$; if this were not the case, then there would be a classical strategy that could perform just as well. (To see this, note that $\alpha_t = \beta_t = \gamma_t$ can be chosen for the columns in which $\gamma_t \neq 0$. Since there is at most one column left, that column can be specified in the classical strategy to match the hyperbit strategy.)

Next, suppose the optimal strategy for Bob is given by the unit vectors $\vec{y}_1$ and $\vec{y}_2$. Recall then the optimal strategy for Alice is determined in Equation~\ref{eq:alice-strat-given-bob} such that the objective function becomes
\begin{equation}\label{eq:3x2-bias}
\begin{split}
I({\bf S}) &= \sum_s \lVert C_{s1}\vec{y}_1+C_{s2}\vec{y}_2 \rVert\\
&= \sum_s\sqrt{C^2_{s1}+C^2_{s2}+2C_{s1}C_{s2}\vec{y}_1\cdot\vec{y}_2}.
\end{split}
\end{equation}
We see then that Bob's strategy is rotationally-invariant, i.e. only the value $\cos\theta\equiv \vec{y}_1\cdot\vec{y}_2$ affects the objective function. It suffices, then, to characterize the angle $\theta\in[0,\pi]$, or the value $z \equiv \cos\theta\in [-1, 1]$. 

This problem is now an optimization problem, seeking to maximize
\begin{equation}\label{eq:f(z)}
\begin{split}
f(z) &= \sqrt{C_{11}^2+C_{12}^2 + 2C_{11}C_{12}z} \\
&+\sqrt{C_{21}^2+C_{22}^2 + 2C_{21}C_{22}z}\\
&+\sqrt{C_{31}^2+C_{32}^2 + 2C_{31}C_{32}z}.
\end{split}
\end{equation}
The concavity of this function with respect to $z$ means that the optimal solution occurs at $z=z^{*}$, according to the following cases:
\begin{enumerate}
	\item If $f'(-1) < 0$, then $f'(z) < 0\,\,\, \forall z\in[-1, 1] \implies z^* = -1$
	\item If $f'(1) > 0$, then $f'(z) > 0\,\,\, \forall z\in[-1, 1]\implies z^* = 1$
	\item Otherwise, $z^{*}\in (-1, 1)$ such that $f'(z^*) = 0$.
\end{enumerate}

Note that in the first two cases, the objective function value is given by
\begin{align}
f(-1) &= \lvert C_{11} - C_{12}\rvert + \lvert C_{21} - C_{22}\rvert + \lvert C_{31} - C_{32}\rvert \\
&= \lvert C_{11} - C_{12}\rvert + \lvert C_{21}\rvert + \lvert C_{22}\rvert + \lvert C_{31} - C_{32}\rvert
\end{align}
and
\begin{align}
f(1) &= \lvert C_{11} + C_{12}\rvert + \lvert C_{21} + C_{22}\rvert + \lvert C_{31} + C_{32}\rvert \\
&= \lvert C_{11} \rvert + \lvert C_{12}\rvert + \lvert C_{21} + C_{22}\rvert + \lvert C_{31}\rvert + \lvert C_{32}\rvert
\end{align}
respectively. Note the simplifications come from the sign matrix we are considering (Equation~\ref{eq:sign-matrix}). 

If the classical strategies (labeled previously as) 2 or 3 are optimal, then $z = 1$ gives a quantum strategy with the same objective function value. When classical strategy 5 is optimal, then $z = -1$ gives a quantum strategy with the same objective function value. Thus, if any of these three classical strategies are optimal, but $z^*$ is found to be in the range $(-1, 1)$, there will exist a quantum advantage.

Note one final analytical feature that is only true in this small case is that the equation $f'(z^*) = 0$ can actually be solved analytically. The exact form is omitted here, but equation can be written in terms of a quartic polynomial in $z^*$, for which there happens to be a closed form formula for the roots. 

\subsection{Larger Games}\label{sec:larger-games}
For larger games, the analysis can proceed in a similar vein as in Section~\ref{sec:3x2-games}; however, the mathematical elegance and existence of closed-form solutions may drop off to the point that numerical methods would be preferred or required. For $M\times 2$ games, with $M>3$, Equation~\ref{eq:3x2-bias} can still be used. The analysis which found quantum advantages in certain regimes that correspond to certain optimal classical strategies can also be repeated.

For $M\times N$ games, with $N>2$, the analysis is less extensible. With more $\vec{y}$ vectors to consider, we can no longer independently parameterize each inner product. At that point, the restrictions essentially become those of an SDP, which again suggests the use of numerical methods instead.

\subsection{Visualization of Game Values and Advantages}\label{sec:visualization}
To aid in understanding the conditions under which advantages can arise, a visual plot would be beneficial. The challenge with plots, however, is that each game is specified by many parameters, in the form of the full ${\bf C}$ matrix. Even the simplest, nontrivial game we considered has $3\times 2 = 6$ degrees of freedom. 

Nevertheless, interesting one-parameter slices over the space of possible games can be considered through a parameterization:
\begin{equation}\label{eq:parameterization}
{\bf C} = {\bf A} + {\bf B}t.
\end{equation}
Here, ${\bf A}$ and ${\bf B}$ are fixed matrices of the same dimension as ${\bf C}$, and $t$ is a numerical parameter to be varied. Sweeps over a range of $t$ can traverse various regimes in which the unlimited, classical, and hyperbit communication cases may categorically vary. 

For concreteness, we consider a specific family of games, given by
\begin{align}
{\bf A} &= \begin{pmatrix}
10 & 1\\
10 & -2\\
-10 & -10
\end{pmatrix}\label{eq:A-parameterization} \\
{\bf B} &= \begin{pmatrix}
0 & 2\\
0 & -1\\
0 & 0
\end{pmatrix}\label{eq:B-parameterization}.
\end{align}
Note the parameterization has been selected carefully, with several properties. First, the sign matrix of ${\bf A}$ is the same as that considered in Equation~\ref{eq:sign-matrix}. Second, the entry with smallest absolute value in ${\bf A}$ is $A_{12}$, which means the smallest absolute value entry for ${\bf C}$ is $C_{12}$ when $t = 0$. As $t$ increases, both $C_{12}$ and $C_{22}$ increase in absolute value; however, the former increases faster, meaning the latter will at some point be the smallest in absolute value. As $t$ increases even more, however, the other four values with absolute value $10$ will become the smallest; for our purposes, it is useful to note that $C_{21}$ is one of these values. The sparsity of ${\bf B}$ was also intentional, to simplify analysis to considering only a sweep varying two parameters of ${\bf C}$. 

\begin{figure*} 
	\includegraphics[width=0.75\textwidth]{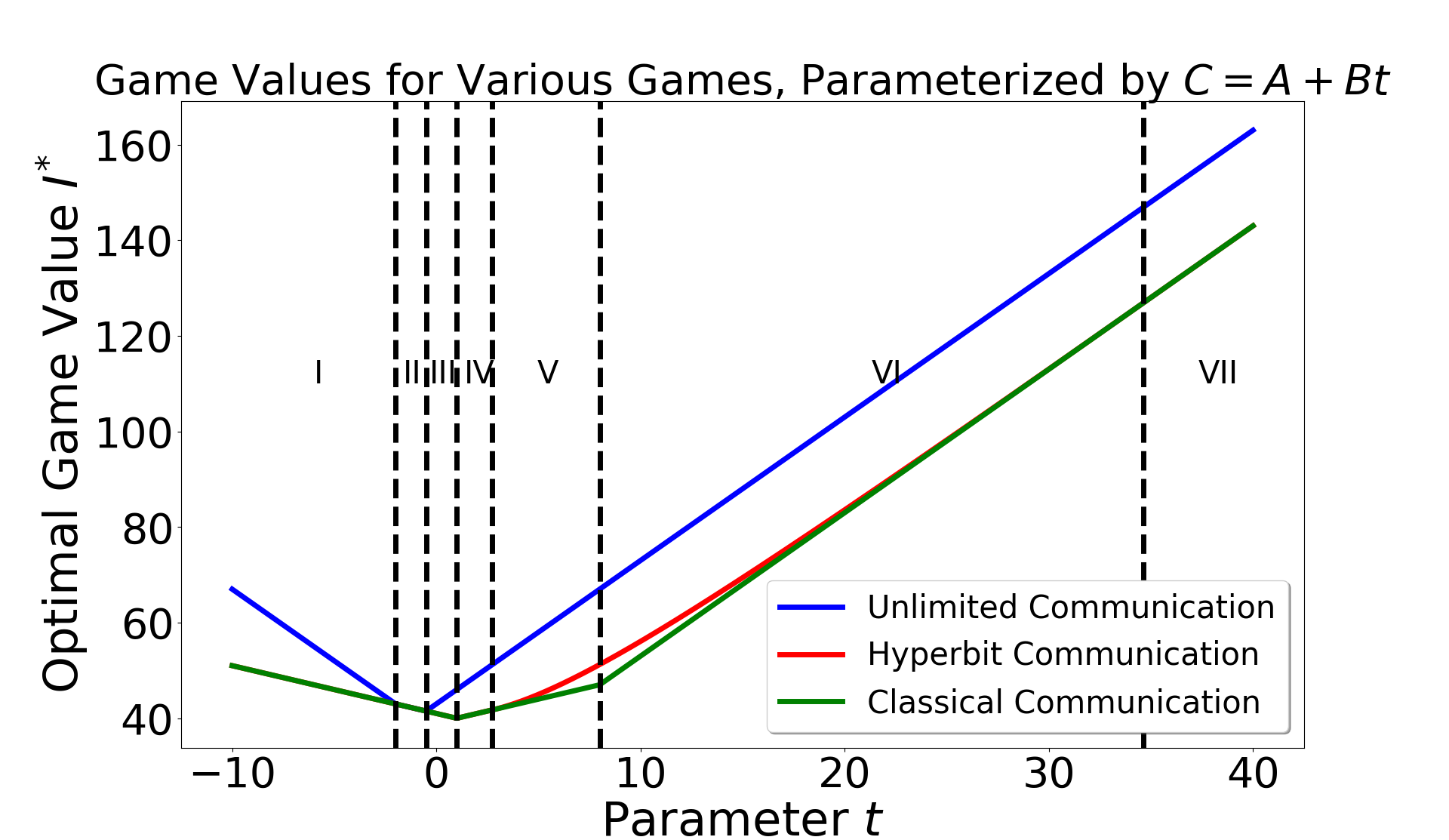}
	\caption{A sweep of the game value for the various strategies considered, with the game parameterized by $C = A + Bt$. Note that seven regions are labeled for categorically different regimes; see the main text for details.}
	\label{fig:sweep}
\end{figure*}

The corresponding sweep for $t\in [-10, 40]$ is given in Figure~\ref{fig:sweep}. It is given with several vertical partitions and labeled regions, the significance of which we will describe.

First, remark that the locations of the vertical lines demarcating the regions are at 
\begin{align}
t_1 &= -2\\
t_2 &= -0.5\\
t_3 &= 1\\
t_4 &= \frac{\sqrt{97}+1}{4}\approx 2.7122\\
t_5 &=8\\
t_6 &= \frac{3\sqrt{601}+65}{4}\approx 34.6365.
\end{align}
Their significance is as follows. As $t$ decreases from $0$, the sign matrix of ${\bf C}$ undergoes two transitions, at $t_1$ and $t_2$, when the signs of $C_{22}$ and $C_{12}$ flip, respectively. As $t$ increases from $0$, $\lvert C_{12}\rvert$ eventually overtakes $\lvert C_{22}\rvert$; this occurs at $t = t_3$, when both absolute values are $3$. For $t_4$ and $t_6$, recall the definition of $f(z)$ given in Equation~\ref{eq:f(z)} and the optimal solution $z^{*}$ conditions described in Section~\ref{sec:3x2-games}. As $t$ continues to increase, there comes a point when $f'(1) < 0$; this occurs for $t\in (t_4, t_6)$. For all $t > t_2$, we have $f'(-1) > 0$. These two conditions amount to the fact that the hyperbit value is maximized for $z^{*}\in(-1, 1)$ when $t\in (t_4, t_6)$. Lastly $t_5$ is the point when $\lvert C_{22}\rvert = 10$; for larger $t$ values, it can be considered that $C_{21}$ is (tied for) the smallest absolute value.

With this understanding, we can now consider the various labeled regimes, and when advantages appear and disappear.

It is most convenient to start with region III. In this region, we have $I_C = I_H < I_U$. The lack of hyperbit advantage is because $f'(1) > 0$ in this region. The presence of the unrestricted communication advantage is due to the nature of the sign matrix of ${\bf C}$. 

Next, consider region IV. The analysis for region III still holds; the only difference is that there is a categorical shift in the optimal classical strategy, due to $ C_{22}$ replacing $C_{12}$ as the smallest absolute value element. 

In region V, the presence of a hyperbit advantage appears. This is due both to the fact that now $f'(1) < 0\implies z^{*}\in (-1,1)$, and also the fact that $C_{22}$ is the smallest absolute value element.

In region VI, the classical strategy again shifts as $C_{22}$ is replaced as the smallest absolute value element. Because $C_{21}$ is (tied for) the smallest absolute value element, and still $z^{*}\in (-1, 1)$, the hyperbit advantage still exists. 

Then, at last in region VII, the hyperbit advantage disappears due to $f'(1)$ again exceeding $0$.

For the regions less than $t_2$, consider how the changes to the sign matrix of ${\bf C}$ affect the game strategies and game values. In region II, the sign matrix becomes
\begin{equation}
\text{sgn } {\bf C}  = \begin{pmatrix}
1 & -1\\
1 & -1\\
-1 & -1
\end{pmatrix}.
\end{equation}
But now that there are only two distinct rows, the the optimal classical strategy matches the optimal unlimited communication strategy. Thus, all advantages disappear and $I_C = I_H = I_Q$. 

Finally, in region I, the sign matrix changes to 
\begin{equation}
\text{sgn } {\bf C}  = \begin{pmatrix}
1 & -1\\
1 & 1\\
-1 & -1
\end{pmatrix}.
\end{equation}
The first two rows can be swapped (as a valid transformation mentioned in Section~\ref{sec:red-and-trans}) to restore the sign matrix at $t = 0$, and the same analysis for region III holds.

The important, high level takeaway is that a one-parameter sweep, as we have plotted here, can quickly show the various regimes one must consider when analyzing games. In some regimes (I, III, IV, VII), there is no hyperbit advantage, despite a difference in classical and unlimited communication game values. In other regimes (II), there is no advantage between any of the three. And, lastly, in some regimes (V, VI) there are advantages between all three; this is the regime of guaranteed quantum advantage. Additionally, the changes in the sign matrix and transitions to different smallest element are apparent from the sweep as well.

\section{Blackjack-Specific Results}\label{sec:backjack-results}
Our games are parameterized based on Bob's initial face-up card, as well as the cards left in the shoe after only faceup cards have been dealt. This fully specifies the probability distribution of the face-down cards $s$ and $t$ to be dealt to Alice and Bob, as well as the probability distribution of the first cards Bob would be dealt if his action was ``hit.''

In cases when the shoe is large, e.g. when a full 52-card deck is left to be dealt, no quantum advantage was found. This does not necessarily rule out advantage, both due to the fact that our quantum strategies may not be general and that our search was not exhaustive. Nevertheless, our analyses has shown a negative result. This makes sense: when there are many cards left in the shoe, the amount of private information Alice has is minimal. That is to say, Alice's face-down card does not significantly affect the probability distribution of the shoe, and therefore will not affect Bob's strategy. In the limit when infinitely many cards are in the deck, the Alice's face down card has no affect at all on Bob's potential strategy. Thus, it is more likely to find quantum advantages in configurations which have only a few cards left in the shoe. The face down card that Alice receives then has a much greater impact on Bob's prospective outcomes and strategy.

Indeed, when the shoe is reduced to just a few cards, advantages do arise. We exhaustively enumerated cases in which, after only the face up cards were dealt, the remaining shoe had between 3 to 8 cards left. There were definitively no advantages found for the case of 3 cards, but advantages were found and enumerated for the cases of 4 to 8 card shoes. 

Many concrete cases correspond to $3\times 3$ dimensional games. Such an example is given as follows. Note first that Alice's face up card can be anything, as it does not affect any players' strategy. We consider a configuration where Bob and the dealer are dealt a 9 and 10 face up, respectively, and the remaining shoe contains two Aces, an 8 and a 10. The advantage amount, calculated from the difference between objective values $I({\bf S})$, is 0.0087.

\begin{figure*}
\begin{quantikz}\lstick[wires=2]{$\ket{0}^{\otimes 2}$\\Alice Qubits} & \gate{H}\gategroup[wires=4,steps
=3,style={dashed,
rounded corners,inner sep=6pt}]{Entangled State Prep} & \ctrl{2}& \qw & \qw & \gate{R_z(\theta_1^A)}\gategroup[wires=4,steps
=10,style={dashed,
rounded corners,inner sep=6pt}]{Hyperbit Measurement} &\qw & \targ{} &\qw& \qw &\qw & \targ{} &\qw& \gate{X} & \meter{}\\
& \gate{H} & \qw & \ctrl{2}& \qw & \qw &\gate{S^{\dagger}} & \ctrl{-1} & \gate{H} & \gate{R_z(\phi_2^A)} &\gate{H} & \ctrl{-1} & \gate{S} &\qw & \qw\\
\lstick[wires=2]{$\ket{0}^{\otimes 2}$\\Bob Qubits} & \qw & \targ{} & \qw & \qw & \gate{R_z(\theta_1^B)} &\qw & \targ{}&\qw& \qw &\qw & \targ{} &\qw& \gate{X} & \meter{}\\
& \qw & \qw& \targ{} & \qw & \qw & \gate{S^{\dagger}} & \ctrl{-1} & \gate{H} & \gate{R_z(\phi_2^B)} &\gate{H} & \ctrl{-1} & \gate{S} &\qw & \qw\end{quantikz}
\caption{Quantum circuit for the optimal hyperbit strategy when Bob and the dealer have face up cards 9 and 10, respectively, and the shoe contains cards [A, A, 8, 10]. The $\theta$ and $\phi$ rotation angles for both Alice and Bob are specified in Table~\ref{tab:blackjack-example-angles} and depend on which facedown card each player is dealt. It is important to note that Alice and Bob's measurements, which are conventionally 0/1, must be converted to -1/+1, and that Bob's action is based on the product of the two measurements.}\label{fig:circuit}
\end{figure*}
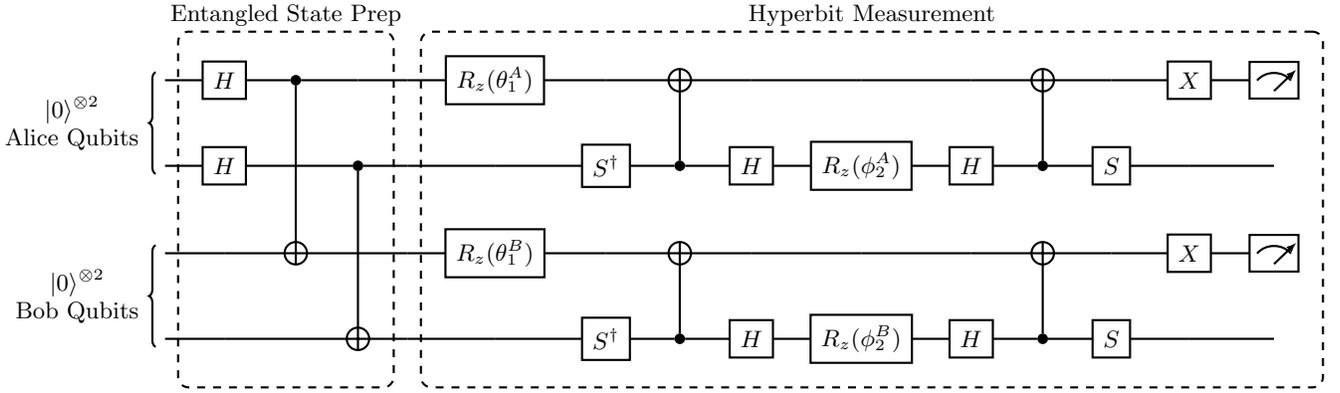

Using the algorithm specified in Section~\ref{sec:hyperbit-in-practice}, we note that the strategy involves each player getting $2$ qubits. We can then specify Alice and Bob's strategy in terms of the rotation angles $\theta_1, \theta_2$ and $\phi_2$ for both players, depending on their face-down card. Note that $\theta_2 = 0$ in all cases for both players, since we only have a $3\times 3$ game. The calculated values for $\theta_1$ and $\phi_2$ are given in Table~\ref{tab:blackjack-example-angles}, and the corresponding circuit is presented in Figure~\ref{fig:circuit}.

\begin{table}
\caption{\label{tab:blackjack-example-angles}
For the example when Bob and the dealer have face up cards 9 and 10, respectively, and the shoe contains [A, A, 8, 10], the tables below specify Alice and Bob's strategy. Note that the measurements Alice and Bob make depend on the face down cards each player is dealt, and that the angles specified correspond to the description from Section~\ref{sec:hyperbit-in-practice}.}
\begin{tabular}{| c || c | c || c | c|}
\hline
Face down card & Alice $\theta_1$ & Alice $\phi_2$ & Bob $\theta_1$ & Bob $\phi_2$\\
\hline
A & 0 & 0 & -2.90 & 1.11e-4\\
8 & 2.99 & 0 & 2.45& 3.95e-4\\
10 & -1.35 & 6.04e-4 & -3.07 & 0 \\
\hline
\end{tabular}
\end{table}

In our search for quantum advantages in small shoe games, a couple of general trends arose. Advantageous configurations tend to have aces left over in the shoe, as they are the lowest risk, highest reward card (they can serve as the highest value of 11, while defaulting to the lowest value of 1 to avoid busting). Furthermore, Bob often starts with a high face up card, leading to scenarios where Bob must weigh the risk of busting. In these cases, Alice's private information would be of great help to Bob, as it would help him weigh the risk and reward of hitting.

\begin{figure}[!th]
\includegraphics[width=0.5\textwidth]{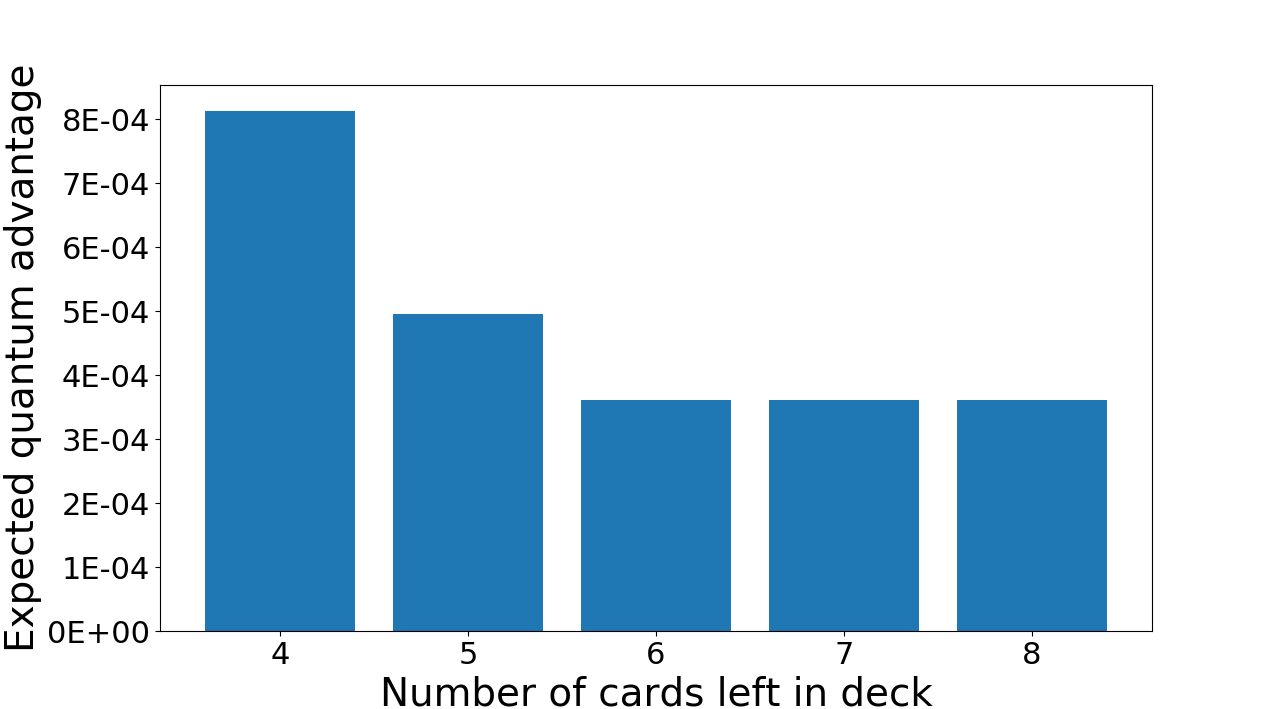}
\caption{The expected advantage amount plotted as a function of the shoe size. Note that the expected advantage seems to fall off and then plateau for larger deck sizes.}
\label{fig:advantage-plot}
\end{figure}

Finally, we also see that advantages tend to be larger and more frequent when fewer cards are left in the deck. This can be seen from Figure~\ref{fig:advantage-plot}, which plots the expected advantage amount across shoe-sizes 4 through 8. Note when we take expectations over shoes of size $k$ (for $k=4,5,6,7,8$) we sample the $k$ cards from an infinite number of full 52-card decks. Text files containing all advantageous configurations for these shoe sizes can be found on our Github repository, which is provided in Appendix~\ref{app:repo}.

\section{Conclusions}
We have provided a broad analysis framework to search for quantum advantages in communication-limited games, using the hyperbit model as our major theoretical tool and semidefinite programming as our major computational tool. This framework was concretely applied to the game of blackjack, and was able to successfully find quantum advantages in certain, small-shoe configurations.
Because of the generality of our framework, a future direction would be to apply it to other games. Any cooperative, multi-player game in which private information can be conveyed in a classical communication-limited can be analyzed using our results. Another direction to continue is to search for more general, yet still computationally tractable, quantum strategies; as mentioned, while our hyperbit model can prove the existence of quantum advantages, the absence of hyperbit advantages cannot completely rule out general advantages. Finally, we hope to see experimental data using our hyperbit strategy algorithm, applied on a small system. As mentioned, each player's strategy can be made so that multi-qubit operations only act on consecutive qubits. This makes the resulting circuit ready for NISQ devices with limited qubit connectivity.

\begin{acknowledgments}
JXL acknowledges support from the MIT Undergraduate Research Opportunities Program (UROP).  AWH was funded by NSF grants CCF-1452616, CCF-1729369, PHY-1818914 and ARO contract W911NF-17-1-0433. AVN was partially supported by NSF grant CCF-1452616. JAF is supported by U.S. Department of Energy Contract DE\verb|-|SC0011091 and NSF award 1505678.  JAF would also like to thank C.~Pollak and his monthly ``Lonely Poker Hearts Club Band" for the topic inspiration.
\end{acknowledgments}

\appendix
\section{General Blackjack Rules}\label{app:blackjack-rules}
As mentioned in Reference~\cite{baldwin1956optimum}, most casinos play blackjack with the same, high level rules, but differ in the specifics. The authors of that paper set out a standardized ruleset, which we deviate slightly from for clarity and ease of analysis. The particular set of rules and conditions we have chosen is described as follows.

Blackjack is played with a set of standard playing card, with each card being either an Ace, a number 2 through 10, inclusive, or a face card (Jack, Queen, or King). All cards begin face-down and unused; this set of unused cards is referred to as the \textit{shoe}.

Among the actors in this game is a single \textit{dealer} and some number of \textit{players}. For our purposes, we will consider just two players, who are named Alice and Bob.

The game begins with the deal. Cards are dealt sequentially, without replacement, from the top of the shoe to an actor either face-up, for all actors to see, or face-down, private to the individual actor. The dealer receives a card face-up from the shoe, while each player receives two cards, one face-up and one face-down. 

At any point, the players and dealer have a particular hand value equal to the sum of the values of the cards in their hand. Each numbered card, from 2 to 10, is worth its numerical value. Each face card (i.e. Jack, Queen, King) is worth 10 points. Lastly, an Ace can be worth either 1 or 11 points, depending on the situation:
\begin{itemize}
	\item  If the Ace can be chosen as 11 points without the player's hand value exceeding 21, then it is chosen to be so.
	\item Otherwise, the Ace is chosen to be worth 1 point.
\end{itemize}
The former kinds of hands are known as \textit{soft} hands. The latter kinds of hands, and hands which do not contain any Aces, are known as \textit{hard} hands.

For each player, the goal of the game is to have a hand value greater than that of the dealer, without exceeding 21. If, at any point, a card holder's hand value exceeds a hard 21 (in particular, the value cannot be made lower by converting any Ace values from 11 to 1), the individual automatically loses. Note that the players only compete with the dealer, and not each other.

Before beginning play, both Alice and Bob make bets. If a player wins the round, the player's payoff equals that of their bet (i.e. they get their bet back, and an additional amount equaling their bet). If they lose the round, their payoff equals the negative of their bet (i.e. they lose their bet). Finally, in the case of a tie, the player simply receives their bet back and their payoff is zero.

Following the bets, play begins with Alice. She has one of two choices: \textit{hit} or \textit{stand}. In the former action, Alice will receive another card, face-down, from the top of the shoe; in the latter, Alice will voluntarily end her turn. Alice can choose to hit as many times as she wishes, unless she \textit{busts}; this happens when her hand value exceeds a hard 21, at which point she automatically loses and is forced to end her turn. Play then proceeds with Bob, who has the same rules as Alice.

If at least one player stands before busting, the round completes with the dealer. The dealer always plays a fixed strategy, depending on if they have a hard or soft hand. The dealer will hit until their hand value reaches or exceeds a hard 17 or a soft 18, at which point they stand~\footnote{Note that this is a rule that most commonly differs from casino to casino. For concreteness, this is the convention we adhere to.}. 

The player
\begin{itemize}
	\item wins if their hand value exceeds that of the dealer, or the dealer busts;
	\item ties if their hand value equals that of the dealer;
	\item and loses if their hand value is less than that of the dealer, or if the player busted.
\end{itemize} 
Note that even if the dealer busts, a player that busted first still loses.

While these rules and conditions are consistent with the overall nature of any typical game of blackjack, it should be noted that the rules chosen have been simplified for the sake of clarity and tractability of analysis (for example, advanced actions like doubling down and splitting are not considered). We do not claim these exact rules necessarily match any standard ruleset or ruleset played in a casino. Nevertheless, the most important rules that typically identify the game as a blackjack game are present in our simplified ruleset.

\section{Software Tools}\label{app:repo}
The code that was used in the computational analysis of this paper can be found at the following GitHub repository: \href{https://github.com/joelin0/quantum-blackjack}{https://github.com/joelin0/quantum-blackjack}. The files contain scripts for computing and comparing the optimal strategies for arbitrary games in all three studied communication regimes, as well as scripts specific to blackjack calculations.

\bibliographystyle{unsrt}
\bibliography{main}
\end{document}